\documentclass[11pt]{article}

\usepackage{graphicx}
\usepackage{amssymb}
\usepackage{float}
\usepackage{chet}
\usepackage{dsfont}
\usepackage{ctable}
\usepackage{tablefootnote}

\def\zz{\mathbb{Z}}
\def\OO{\mathcal{O}}

\newcommand{\<}{\langle}
\renewcommand{\>}{\rangle}
\newcommand{\be}{\begin{eqnarray}}
\newcommand{\ee}{\end{eqnarray}}
\def\nn{\nonumber}

\date{July 2018}

\title{Bootstrapping the Minimal 3D SCFT}

\author{Alexander Atanasov$^{a}$, Aaron Hillman$^{a}$, David Poland$^{a,b}$
\emails{(\href{mailto:alex.atanasov@yale.edu}{alex.atanasov},
\href{mailto:aaron.hillman@yale.edu}{aaron.hillman},
\href{mailto:david.poland@yale.edu}{david.poland})@yale.edu}}

\affiliation{$^{a}$Department of Physics, Yale University, New Haven, CT 06520, USA \\
$^{b}$Walter Burke Institute for Theoretical Physics, Caltech, Pasadena, CA 91125, USA}

\abstract{We study the conformal bootstrap constraints for 3D conformal field theories with a $\mathbb Z_2$ or parity symmetry, assuming a single relevant scalar operator $\epsilon$ that is invariant under the symmetry. When there is additionally a single relevant odd scalar $\sigma$, we map out the allowed space of dimensions and three-point couplings of such ``Ising-like" CFTs. If we allow a second relevant odd scalar $\sigma'$, we identify a feature in the allowed space compatible with 3D $\mathcal{N}=1$ superconformal symmetry and conjecture that it corresponds to the minimal $\mathcal{N}=1$ supersymmetric extension of the Ising CFT. This model has appeared in previous numerical bootstrap studies, as well as in proposals for emergent supersymmetry on the boundaries of topological phases of matter. Adding further constraints from 3D $\mathcal{N}=1$ superconformal symmetry, we isolate this theory and use the numerical bootstrap to compute the leading scaling dimensions $\Delta_{\sigma} = \Delta_{\epsilon} - 1 = .58444(22)$ and three-point couplings $\lambda_{\sigma\sigma\epsilon} = 1.0721(2)$ and $\lambda_{\epsilon\epsilon\epsilon} = 1.67(1)$. We additionally place bounds on the central charge and use the extremal functional method to estimate the dimensions of the next several operators in the spectrum. Based on our results we observe the possible exact relation $\lambda_{\epsilon\epsilon\epsilon}/\lambda_{\sigma\sigma\epsilon} = \tan(1)$.}

\begin{document}

\maketitle
\toc

\newsec{Introduction}

The conformal bootstrap has emerged as a powerful nonperturbative tool for studying
conformal field theories (CFTs) in $D>2$, with numerous applications ranging 
from critical phenomena to holography. It has enabled us to make precise 
quantitative predictions for strongly-interacting CFTs, such as the 
3D Ising and $O(N)$ models~\cite{ElShowk:2012ht,Kos:2013tga,El-Showk:2014dwa,Kos:2014bka,Kos:2015mba,Kos:2016ysd}, along with numerous other CFTs. 
For a recent review of these developments, see~\cite{Poland:2016chs,Poland:2018epd}.

In this work we will apply conformal bootstrap techniques to critical CFTs with a $\zz_2$ or 
parity symmetry using mixed scalar correlators, extending the bounds of~\cite{Kos:2016ysd} 
away from the 3D Ising critical point. First in Sec.~\ref{sec:mixed} we focus on the allowed region of CFTs with two 
relevant operators $\sigma$ and $\epsilon$, extending the map~\cite{Kos:2014bka,Kos:2016ysd} of the allowed values of the scaling dimensions and OPE coefficients. We observe that the tip of the bulk allowed region is close to satisfying the constraints imposed by supersymmetry, and that it can become precisely compatible if another relevant $\zz_2$ or parity-odd scalar operator $\sigma'$ is allowed in the spectrum. A similar story appeared previously in the fermion bootstrap analysis of~\cite{Iliesiu:2015qra,Iliesiu:2017nrv}, where it was conjectured that the supersymmetric theory with two relevant parity-odd scalars coincided with the minimal $\mathcal{N}=1$ superconformal extension of the 3D Ising model. This model has been proposed to be realizable on the boundaries of topological materials~\cite{Grover:2013rc} and has recently seen attention using other techniques such as the $\epsilon$-expansion~\cite{Fei:2016sgs,Zerf:2017zqi,Ihrig:2018hho}.

We show in Sec.~\ref{sec:SCFT} that this theory is restricted to a very small region for relevant values of $\Delta_{\sigma'}$ by combining the constraints of supersymmetry with the input that the super-descendant $\epsilon'$ of dimension $\Delta_{\epsilon'} = \Delta_{\sigma'} + 1$ is the second parity-even scalar. This latter assumption is well-motivated from previous studies in the $\epsilon$-expansion and we will find that a consistent picture of the spectrum emerges. In order to incorporate constraints from 3D $\mathcal{N}=1$ superconformal symmetry, in Appendix~\ref{app:3point} we use the formalism of~\cite{Park:1999cw} to derive the relation between $\lambda_{\sigma\sigma\epsilon}$ and $\lambda_{\epsilon\epsilon\epsilon}$, which has a highly nontrivial intersection with the regions allowed by the bootstrap and restricts $\Delta_{\sigma}$ to a small range. Over this range we study the minimum and maximum values of the OPE coefficient $\lambda_{\sigma\sigma\epsilon}$, the lower bound on the central charge, and in Sec.~\ref{sec:spectrum} we look at the resulting extremal functionals to obtain an approximate picture of the higher spectrum. We conclude in Sec.~\ref{sec:conclusion} by pointing out the compatibility of our results with the possible relation $\lambda_{\epsilon\epsilon\epsilon}/\lambda_{\sigma\sigma\epsilon} = \tan(1)$.

{\bf Note Added:} As this work was being completed, the nice paper~\cite{Rong:2018okz} appeared which has partial overlap with our results of Sec.~\ref{sec:SCFT}. They improve our estimates of $\Delta_{\sigma}$ and $\Delta_{\sigma'}$ in the $\mathcal{N}=1$ Ising SCFT by deriving and incorporating the full SUSY conformal block structure. In addition to our map of the Ising-like CFT bulk region in Sec.~\ref{sec:mixed}, our analysis has added new constraints on the $\mathcal{N}=1$ Ising SCFT OPE coefficients, estimates of the higher spectrum, and a possible exact formula for $\Delta_{\sigma}$.

\newsec{Mixed-Correlator Studies in the Space of 3D Ising-like CFTs}
\label{sec:mixed}

We begin by first considering the space of unitary 3D CFTs with a $\mathbb Z_2$ symmetry and two relevant scalars. We require one of these scalars to be $\mathbb Z_2$-odd and denote it by $\sigma$, while the other one is $\mathbb Z_2$-even and is denoted by $\epsilon$. The pair of scaling dimensions $(\Delta_\sigma, \Delta_\epsilon)$ form a two-dimensional space that we can explore. Unitarity of the theory constrains $\Delta_\sigma$ and $\Delta_\epsilon$ to each be greater than $1/2$, while the operators' relevance requires their scaling dimensions to also be less than $3$. This space has been studied in detail in \cite{ElShowk:2012ht} and \cite{El-Showk:2014dwa} using single-correlator analysis and $c$-minimization techniques. Mixed-correlator studies on this space have been performed in \cite{Kos:2014bka,Kos:2016ysd}. Using a technique involving scanning over the ratio of the leading three-point function coefficients $\lambda_{\epsilon\epsilon\epsilon}/\lambda_{\sigma\sigma\epsilon}$, the authors of \cite{Kos:2016ysd} obtained particularly sharp bounds on the isolated coordinates of the point corresponding to the 3D Ising universality class:
\begin{equation}
	\Delta_\sigma = 0.5181489(10), \qquad \Delta_\epsilon = 1.412625(10).
\end{equation}
Here, we first aim to apply similar techniques to study the larger space of possible ``Ising-like" CFTs away from the 3D Ising point. One reason to do this is to test the conjecture that the Ising CFT is the only CFT in this space. We will not prove this conjecture, but we will place stronger restrictions on the allowed space of scaling dimensions of other ``Ising-like" CFTs.

The OPE structure of the two relevant operators can be schematically written as
\begin{equation}\label{eq:3D_ising_ope}
	\begin{aligned}
		\sigma \times \sigma &\sim \sum_{\OO^+} \lambda_{\sigma \sigma \OO} \OO,\\
		\sigma \times \epsilon &\sim \sum_{\OO^-} \lambda_{\sigma \epsilon \OO} \OO,\\
		\epsilon \times \epsilon &\sim \sum_{\OO^+} \lambda_{\epsilon \epsilon \OO} \OO.
	\end{aligned}
\end{equation}
In the above equations, $\OO^+$ runs over $\zz_2$-even operators of even spin and $\OO^-$ runs over $\zz_2$-odd operators of any non-negative integer spin.

\begin{figure}[t]
  \begin{center}
    \includegraphics[scale=0.7]{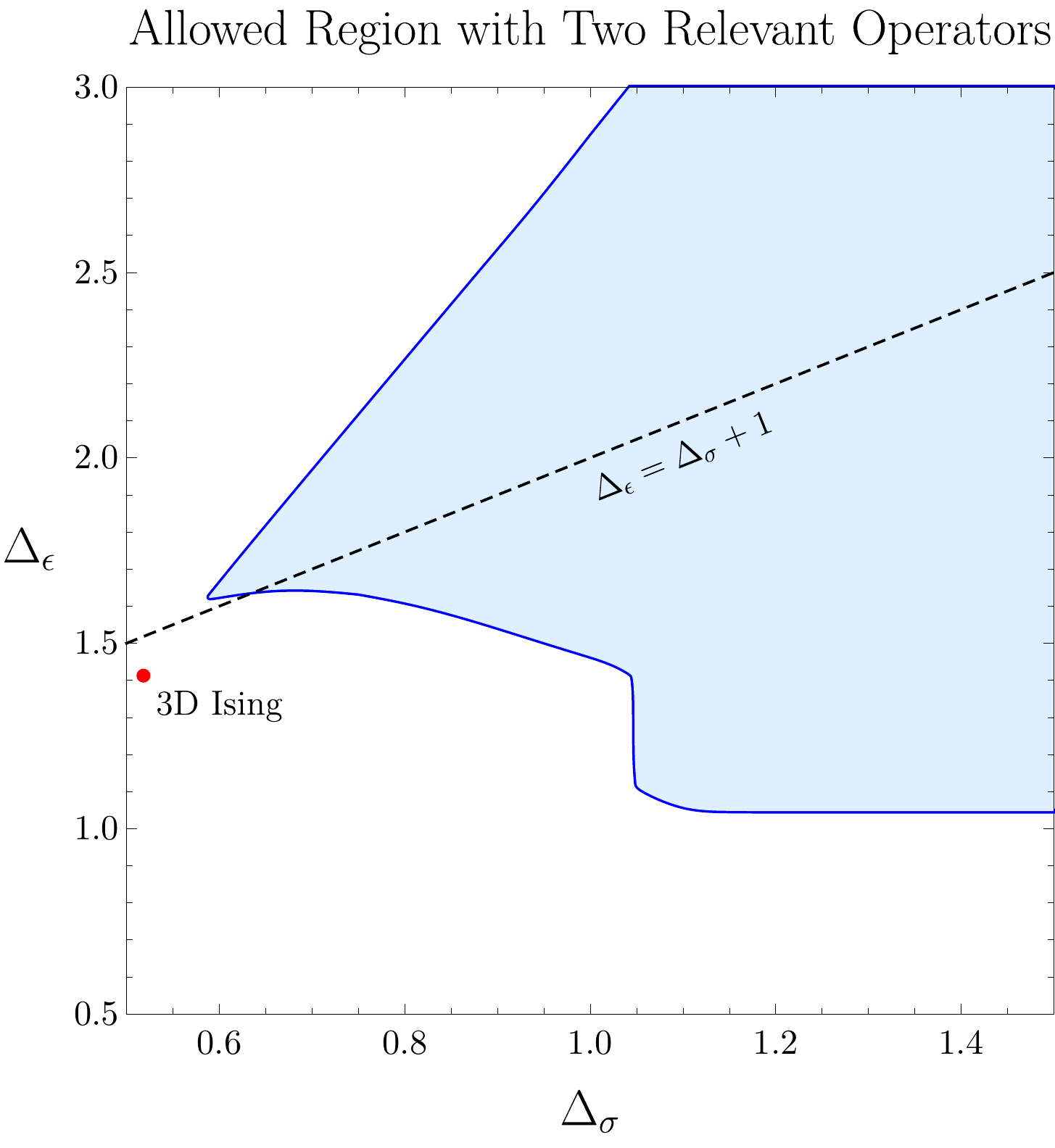}
  \end{center}
  \vspace{-12pt}
  \caption{A plot of the allowed region of 3D CFTs with a $\zz_2$ symmetry. Here we assume two relevant operators, $\sigma$ and $\epsilon$, taken to be $\mathbb Z_2$-odd and $\mathbb Z_2$-even symmetric, respectively. This plot assumes the permutation symmetry $\lambda_{\sigma\epsilon\sigma} = \lambda_{\sigma\sigma\epsilon}$. The shaded area is not excluded. In this plot we use $\Lambda=25$.}
  \label{fig:larger_region}
\end{figure}

The three-point function coefficients additionally satisfy permutation symmetry, e.g.
\begin{equation}\label{eq:3_pt_symmetry}
	\lambda_{\sigma \epsilon \sigma} = \lambda_{\sigma \sigma \epsilon}.
\end{equation}
Here, we incorporate this as a constraint in our formulation of the bootstrap problem for the correlators $\{\<\sigma\sigma\sigma\sigma\>, \<\sigma\sigma\epsilon\epsilon\>, \<\epsilon\epsilon\epsilon\epsilon\>\}$, as described in detail in~\cite{Kos:2016ysd}. The permutation constraint weakens the conditions that a functional must satisfy in order to violate the crossing symmetry bounds, thus excluding a larger region of possible CFTs away from the 3D Ising model as compared with~\cite{Kos:2014bka}. The resulting non-excluded region is shown in Fig.~\ref{fig:larger_region} using derivative order $\Lambda=25$ in the notation of~\cite{Kos:2016ysd}. We compute our bounds using the semidefinite program solver SDPB~\cite{Simmons-Duffin:2015qma} and use cboot~\cite{CBoot} to set up the problem. The various parameters going into the optimization problem are described in Appendix~\ref{app:SDPB}. Going to higher orders of $\Lambda$ may slightly shrink this region, but will likely not eliminate it entirely. 

\begin{figure}[t]
	\begin{center}
		\includegraphics[scale=0.7]{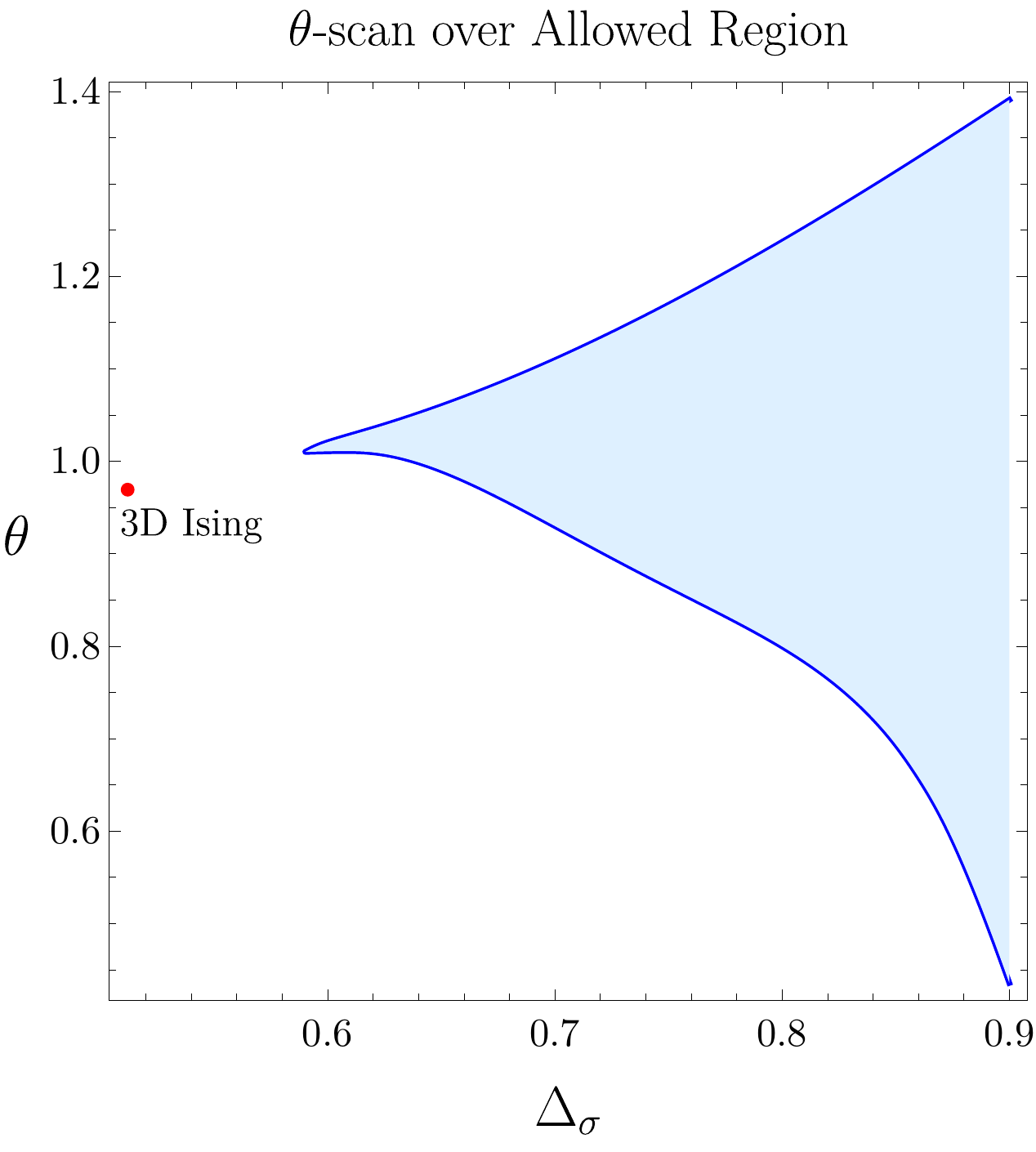}
	\end{center}
	\vspace{-12pt}
	  \caption{A plot showing how the allowed range of $\theta$, taken over all $\Delta_\epsilon$, grows on the non-excluded region of CFTs. In this plot we use $\Lambda=25$.}
	  \label{fig:larger_region_theta}
\end{figure}

Let us make a couple of observations about this allowed region. First, $\Delta_{\epsilon}$ is required to be larger than $1$ over the whole region we have explored. Second, the left-most cusp of the bulk region is close to (but narrowly misses) the line $\Delta_{\epsilon} = \Delta_{\sigma}+1$ that would be required in a supersymmetric theory if $\sigma$ and $\epsilon$ are in the same supersymmetry multiplet. In the next section we will study the possibility that allowing a second relevant odd operator $\sigma'$ will make this feature compatible with supersymmetry.

Before we proceed, it is worth noting here that scanning over the ratios of three-point coefficients $\tan \theta \equiv \lambda_{\epsilon \epsilon \epsilon}/\lambda_{\sigma \sigma \epsilon}$ as implemented in \cite{Kos:2016ysd} does not significantly shrink the size of this larger region as it did for the 3D Ising island. On the other hand, since it turns out to be within our computational capacity to perform a $\theta$ scan over the non-excluded region, we can explore how the range of allowed $\theta$ depends on a given point $(\Delta_\sigma, \Delta_\epsilon)$ in the region. In Figure~\ref{fig:larger_region_theta}, we show the projection of the resulting 3D region to the $(\Delta_\sigma, \theta)$ space to give a sense of how the $\theta$ range grows as $\Delta_\sigma$ increases. Near the cusp of the region, we can see that the allowed range $\theta$ lives in a small window. The value of $(\Delta_\sigma,  \theta)$ found for the 3D Ising model in \cite{Kos:2016ysd} is additionally plotted in red.

\newsec{The $\mathcal N=1$ Ising SCFT}
\label{sec:SCFT}
In the previous section, we explored the space of 3D CFTs with the same symmetries and relevant operator structure as the 3D Ising CFT. This space has an allowed bulk region away from the Ising island depicted in Figure~\ref{fig:larger_region}.  This bulk region has a kink which narrowly fails to intersect the line $\Delta_\epsilon = \Delta_\sigma + 1$, a constraint on the scaling dimensions of $\sigma$ and $\epsilon$ that holds if these operators are presumed to be in a supersymmetry multiplet.  

In fact, the scaling dimensions in the vicinity of this kink are close to values estimated for the $\mathcal{N} = 1$ SCFT corresponding to the minimal supersymmetric extension of the Ising model. Previous estimates include $\Delta_{\sigma} > .565$ from the conformal bootstrap for single scalar correlators~\cite{Bashkirov:2013vya}, $\Delta_{\sigma} \in (.571,.586)$ from the same system with additional gap assumptions~\cite{Li:2017kck}, and $\Delta_{\sigma} \approx .582$ using the conformal bootstrap for fermions~\cite{Iliesiu:2015qra}. Interpolations between the 2D tri-critical Ising model and the $(4-\epsilon)$-expansion have additionally yielded the estimates $\Delta_{\sigma} \approx .588$~\cite{Fei:2016sgs} and $\Delta_{\sigma} \approx .585$~\cite{Zerf:2017zqi}. Finally, the recent Borel resummation analysis of~\cite{Ihrig:2018hho} gave the fairly precise result $\Delta_{\sigma} = .58365(135)$. Our bootstrap result will be perfectly compatible with this determination.

This 3D SCFT is thought to emerge as an IR fixed point of the theory of a real scalar $\sigma$ and Majorana fermion $\psi$, with interactions $\mathcal{L}_{\text{Int}} = -\frac{g}{2} \sigma \bar{\psi} \psi -\frac{1}{8} (g\sigma^2 + h)^2$ and $h$ tuned appropriately. Equivalently, it can be described in terms of a real superfield $\Sigma = \sigma + \theta \psi + \frac{i}{2} \theta^2 F$ with superpotential $W = h \Sigma + \frac{g}{3} \Sigma^3$. This theory has a parity symmetry under which $\Sigma$ is odd, which will play the role of the $\zz_2$ symmetry described above.

We are then motivated to ask whether the kink may intersect the SUSY constraint line and correspond to the minimal 3D $\mathcal{N}=1$ SCFT if we relax some assumptions about the spectrum.  Instead of examining CFTs with two relevant scalars, we can lower the gap assumption for the next-to-lowest lying $\mathbb Z_2$ odd operator $\sigma'$.  This is also motivated by the extrapolated values of this operator in~\cite{Fei:2016sgs}, which estimated $\Delta_{\sigma'} \approx 2.79$, and the fermion bootstrap analysis in~\cite{Iliesiu:2015qra}, which also proposed $\Delta_{\sigma'} < 3$.

In addition, from supersymmetry we expect a parity-even operator $\epsilon'$ of dimension $\Delta_{\sigma'}+1$ in the same SUSY multiplet as $\sigma'$, assuming $\sigma'$ is the lowest component of the multiplet. Based on the picture of the spectrum obtained from interpolating between 2D and 4D~\cite{Fei:2016sgs}, this is almost certainly the case, since there are no operators of lower dimension that could plausibly be in its multiplet. To obtain tighter constraints, we will further make the assumption that this $\epsilon'$ of dimension $\Delta_{\sigma'} + 1$ is the next parity-even scalar in the spectrum. In principle one could imagine that the next parity-even scalar is in a different kind of multiplet, either as the superconformal primary or in a multiplet where the superconformal primary is a fermion. However, in the $(4-\epsilon)$-expansion one sees that the next parity-even scalar (a mixture of $\sigma^4$ and $\sigma \bar{\psi} \psi$) has dimension $\Delta_{\epsilon'} = 4-\frac{3}{7} \epsilon^2 + \ldots = \Delta_{\sigma'} + 1$, while the other eigenvalue is $\Delta_{\epsilon''} = 4 + \frac{6}{7} \epsilon + \ldots$ (see~\cite{Fei:2016sgs}). Our working assumption is that this operator ordering continues to hold in 3D. The extrapolated results for the correction-to-scaling exponent $\omega$ in~\cite{Zerf:2017zqi} further yield the estimate $\Delta_{\epsilon'} = 3 + \omega \approx 3.84$, which is around the correct dimension to be in a supermultiplet with $\sigma'$ as estimated above; we take this as further guidance that our interpretation is correct.

\begin{figure}[t]
  \begin{center}
    \includegraphics[scale=0.7]{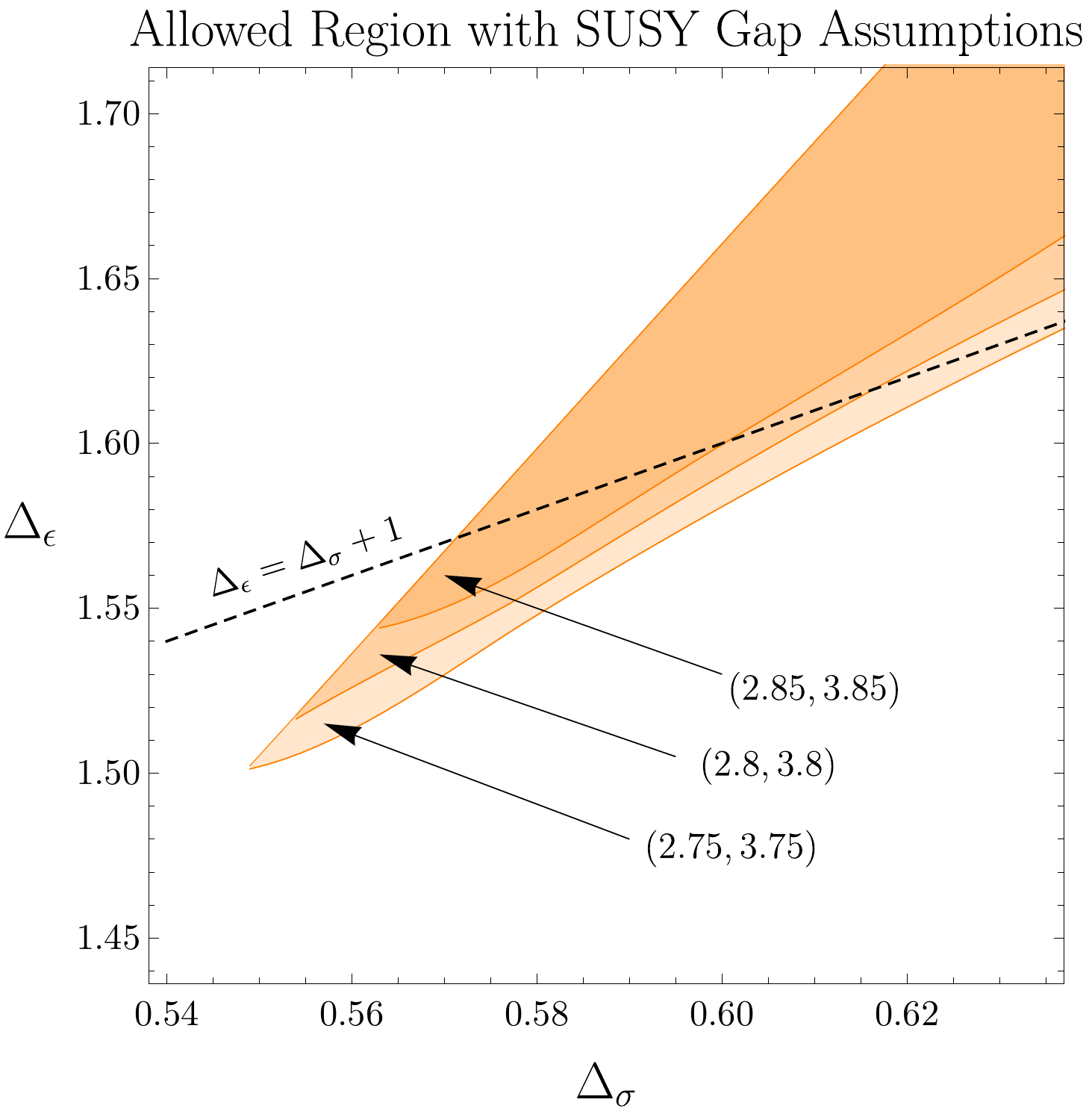}
  \end{center}
  \vspace{-12pt}
  \caption{The shaded areas show the non-excluded regions at various gap assumptions. Here we assume a third relevant parity-odd operator $\sigma'$, together with the relevant operators $\sigma$ and $\epsilon$ from before. We additionally assume that there are no other parity-even scalar operators below $\Delta_{\epsilon'} = \Delta_{\sigma'} + 1$. In this plot we use $\Lambda=25$.}
  \label{fig:susy_line_regions}
\end{figure}

To begin, we present the intersection of the allowed region with the SUSY constraint line for three sets of gap assumptions in Figure~\ref{fig:susy_line_regions}. In each case one can see a clear upper and lower bound on $\Delta_\sigma = \Delta_\epsilon -1$, with the window shrinking as the gap increases.

\begin{figure}[t]
  \begin{center}
    \includegraphics[scale=0.54]{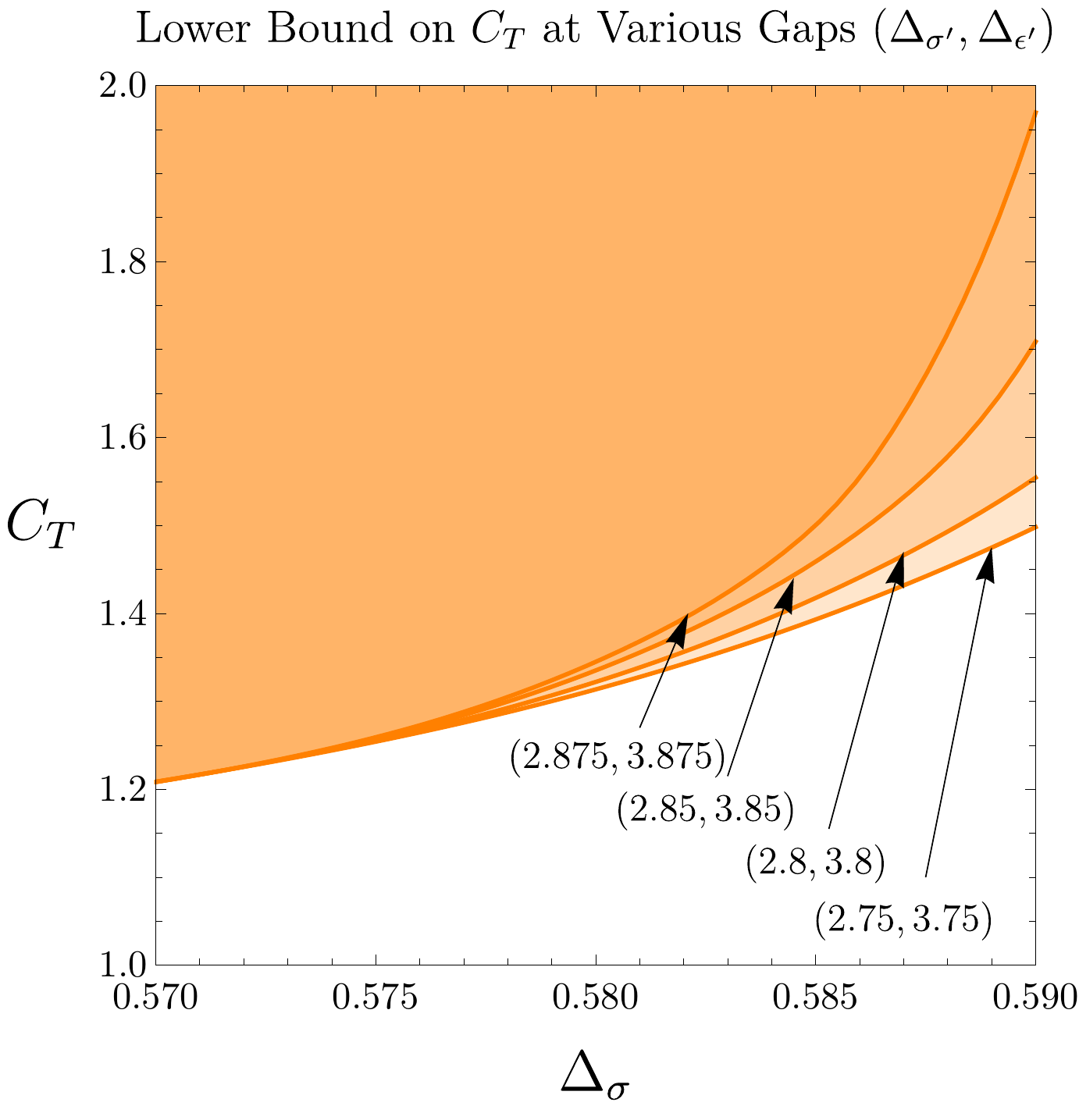} \hspace{45pt} \includegraphics[scale=0.531]{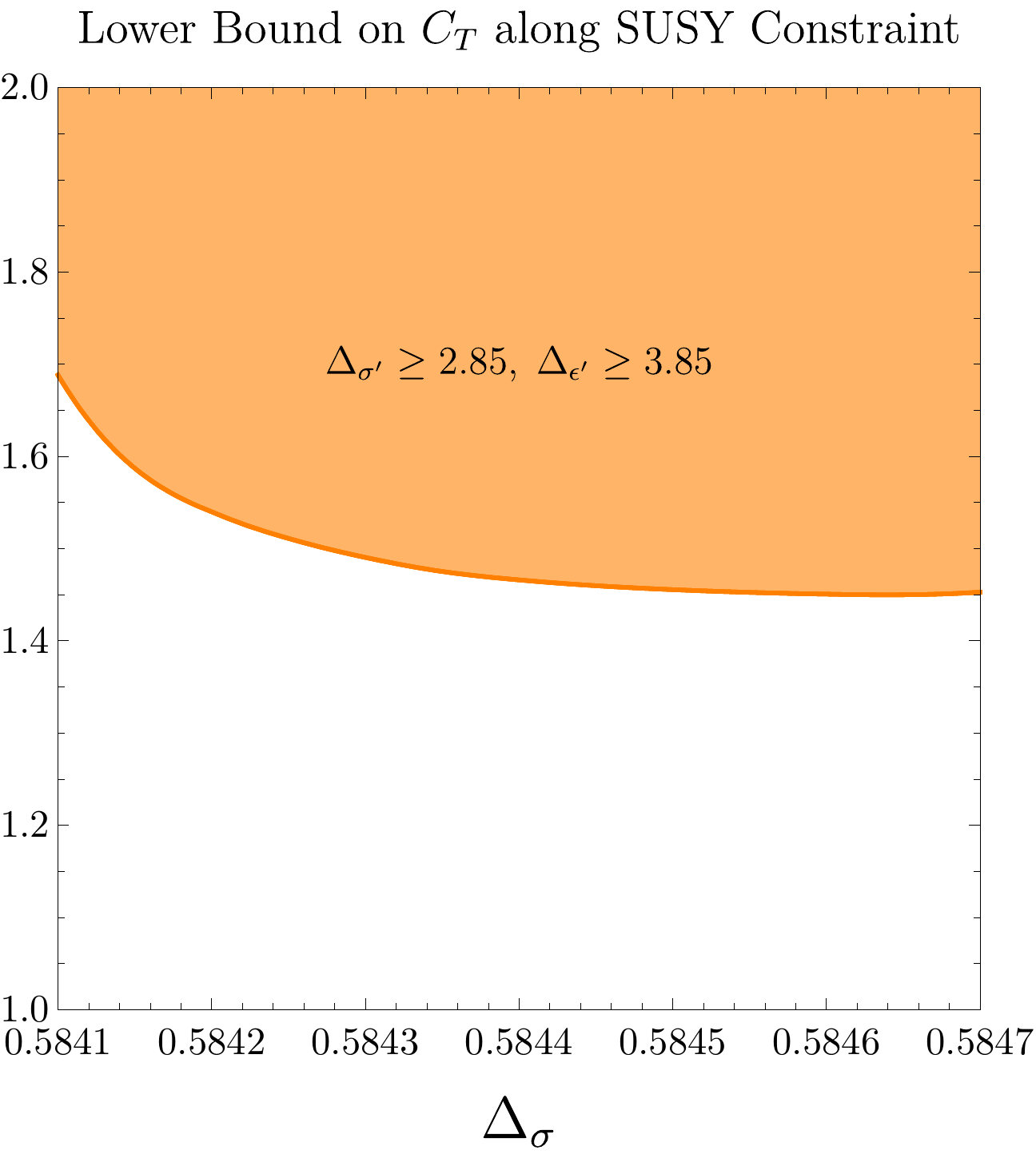}
  \end{center}
  \vspace{-12pt}
  \caption{(Left) A plot of the lower bound on the central charge $C_T$ versus $\Delta_\sigma$, constrained to the SUSY line $ \Delta_\sigma = \Delta_\epsilon - 1$ at gaps $\Delta_{\sigma'} = \Delta_{\epsilon'} -1 \geq \{2.75, 2.80, 2.85, 2.875 \}$. (Right) The lower bound of $C_T$ along the additional constraint of Equation~\eqref{eq:susy}. In these plots we use $\Lambda=25$.}
  \label{fig:cT}
\end{figure}

Another simple plot that we can make using the numerical bootstrap is a lower bound on the central charge $\<TT\> \propto C_T$ along the line $\Delta_{\sigma} = \Delta_{\epsilon}-1$, assuming various values of the $\Delta_{\sigma'} = \Delta_{\epsilon'}-1$ gap. This bound is shown in the left plot of Fig.~\ref{fig:cT}, where $C_T$ is normalized to the value of a free scalar field. A robust conclusion is that $C_T > 1.2$, and at larger values of $\Delta_{\sigma}$ and $\Delta_{\sigma'}$, $C_T$ is forced to even larger values (the right plot will be discussed further below).

\begin{figure}[t]
  \begin{center}
    \includegraphics[scale=0.7]{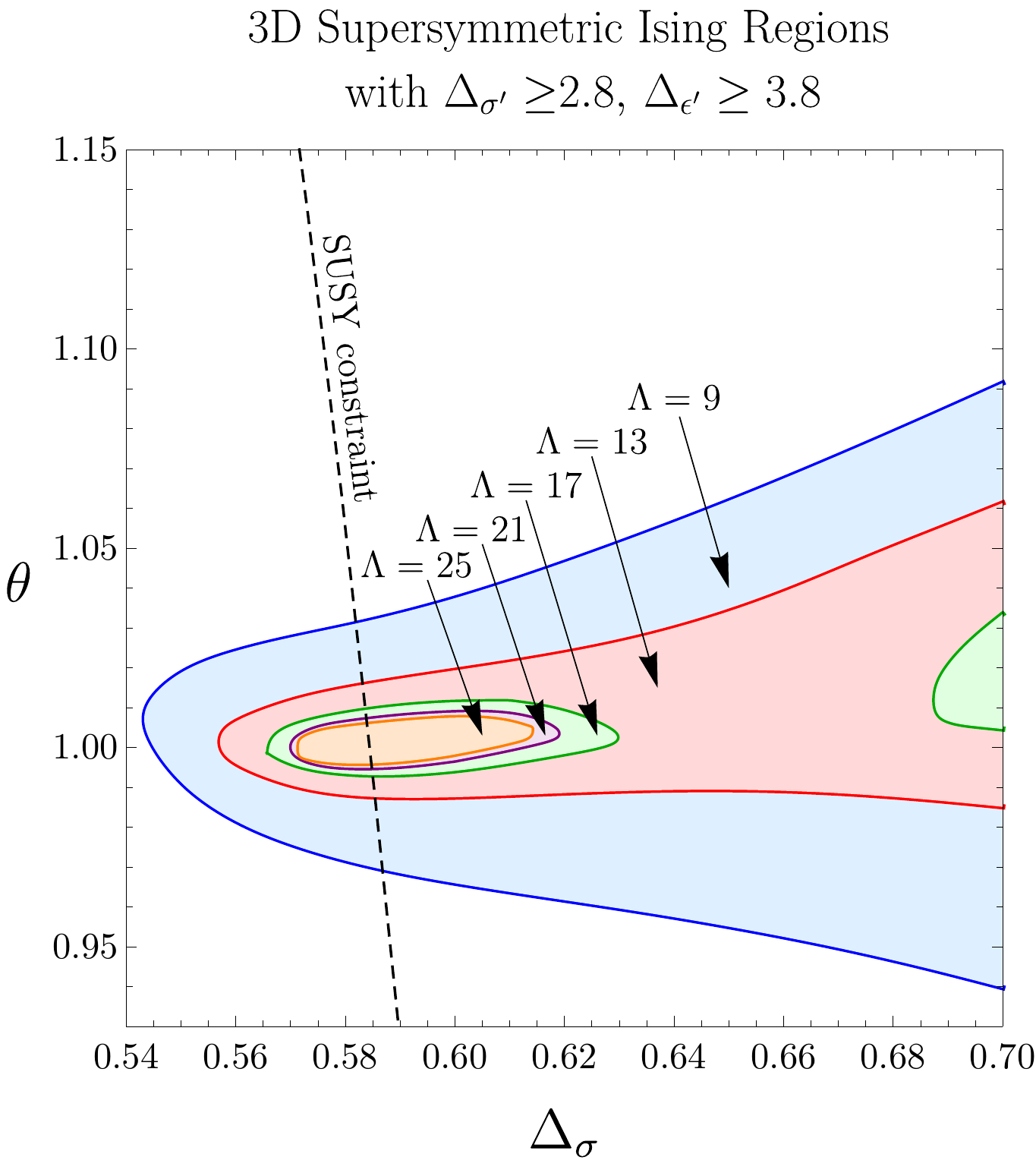}
  \end{center}
  \vspace{-12pt}
  \caption{A plot of the non-excluded regions with gap assumptions of $\Delta_{\sigma'} = \Delta_{\epsilon'}-1 \geq 2.8$ along the supersymmetric line $\Delta_\epsilon = \Delta_\sigma + 1$. The shaded regions are obtained at derivative order $\Lambda = \{9, 13, 17, 21, 25\}$ and denote the non-excluded points.}
  \label{fig:lambda_dep}
\end{figure}

\begin{figure}[t]
  \begin{center}
    \includegraphics[scale=0.7]{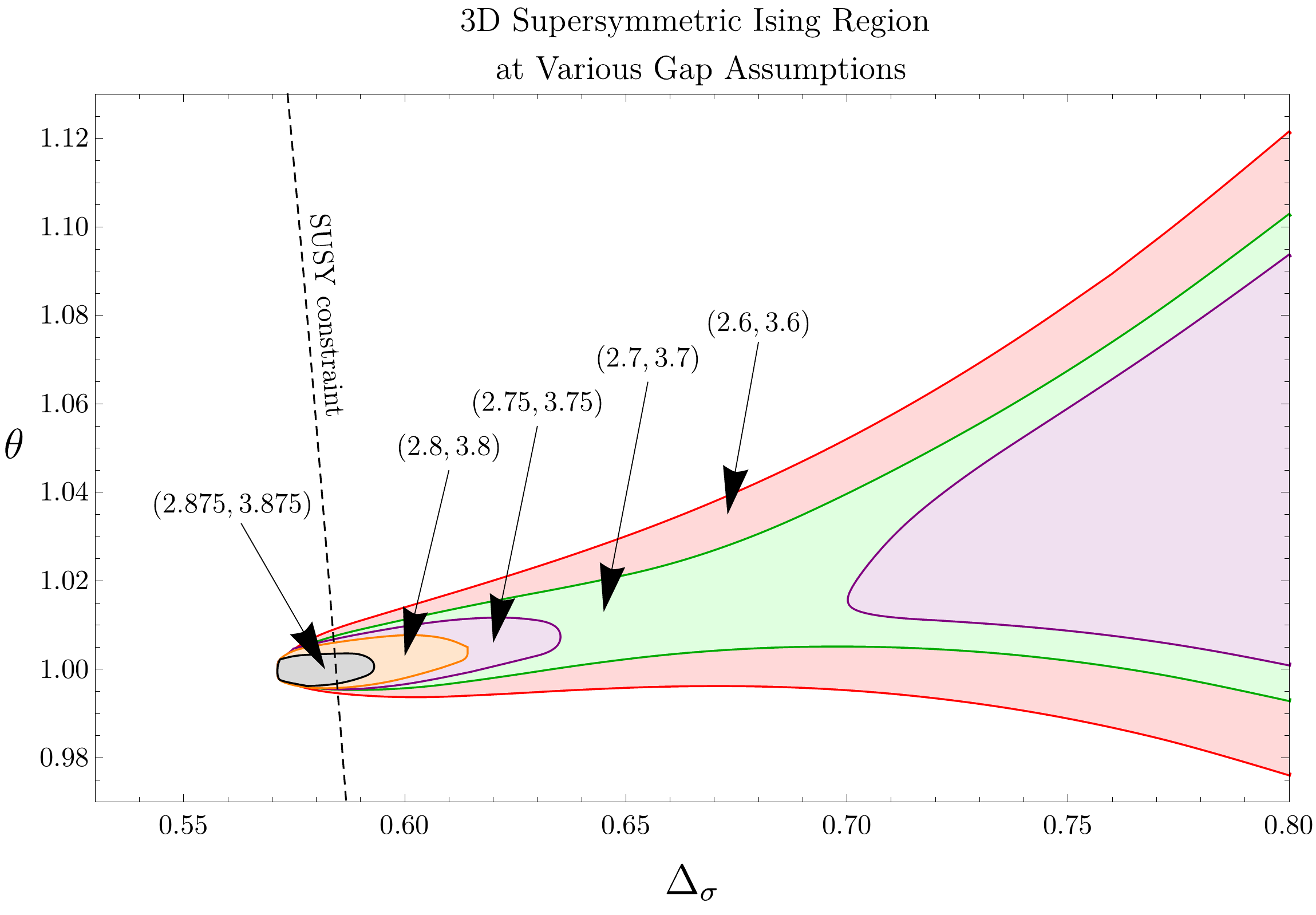}
  \end{center}
  \vspace{-12pt}
  \caption{A plot of the non-excluded regions at various gap assumptions in $(\Delta_\sigma, \theta)$ space along the supersymmetric line $\Delta_\epsilon = \Delta_\sigma + 1$. Again, we assume a second relevant parity-odd operator $\sigma'$ and study different values of the gap until $\Delta_{\sigma'} = \Delta_{\epsilon'}-1$. The shaded areas were obtained at derivative order $\Lambda=25$ and denote the non-excluded points.}
  \label{fig:different_gaps}
\end{figure}

 We can go further by considering the leading three-point function coefficients $\lambda_{\sigma\sigma\epsilon}$ and $\lambda_{\epsilon\epsilon\epsilon}$. First, by treating these coefficients as independent parameters, one can scan over the allowed values of the ratio $\tan(\theta) = \lambda_{\epsilon\epsilon\epsilon}/\lambda_{\sigma\sigma\epsilon}$ as described in~\cite{Kos:2016ysd}. Combined with an assumed gap for the $\sigma'$ multiplet yields closed regions in $(\Delta_{\sigma},\theta)$ space shown in Figs.~\ref{fig:lambda_dep} and~\ref{fig:different_gaps}. The first plot illustrates the effect of increasing $\Lambda$, showing that our regions appear to be relatively converged. The second plot shows how the island shrinks as the gap until the $\sigma'$ multiplet is increased. 
 
 An additional constraint that we have not yet used is the relation between $\lambda_{\sigma\sigma\epsilon}$ and $\lambda_{\epsilon\epsilon\epsilon}$ imposed by $\mathcal{N}=1$ superconformal symmetry. In Appendix~\ref{app:3point} we derive this constraint, given by
 \be\label{eq:susy}
\frac{\lambda_{\epsilon\epsilon\epsilon}}{\lambda_{\sigma\sigma\epsilon}} &=& \frac{3(\Delta_{\sigma}-1)(3\Delta_{\sigma}-2)}{4\Delta_{\sigma}(\Delta_{\sigma}-1/2)},
 \ee
 and shown by the dashed lines in Figs.~\ref{fig:lambda_dep} and~\ref{fig:different_gaps}. It can be seen that this constraint is beautifully compatible with the SCFT living in these islands. Further, the constraint restricts $\Delta_{\sigma}$ to live in a significantly smaller window, even at modest values of the $\sigma'$ gap. On the other hand, we see that the $\sigma'$ gap can be pushed up to $\Delta_{\sigma'} \lesssim 2.9$ and still be compatible with this constraint.\footnote{A gap at precisely 2.9 seems likely to be excluded at higher $\Lambda$, so we have opted to probe gaps up to 2.875 which appears safer upon increasing the derivative order.}  A summary of the minimal and maximal values of $\Delta_{\sigma}$ at different values of the gap will be presented in Table~\ref{tab:spectrum} below.

\newsec{Results and Spectrum Extraction}
\label{sec:spectrum}
 
We can extract additional CFT data for low-lying operators in the spectrum using the ``extremal functional" approach introduced in~\cite{Poland:2010wg,ElShowk:2012hu}. In order to do this we will find the primal-dual optimal functionals placing either upper or lower bounds on the coefficient $\lambda_{\sigma\sigma\epsilon}$.  This can be performed over a set of points in the allowed regions in Figure~\ref{fig:susy_line_regions} and on the SUSY constraint line for a range of gap assumptions.   We find that the minimal and maximal values of $\lambda_{\sigma\sigma\epsilon}$ are attained at the minimal and maximal values of $\Delta_{\sigma}$. The allowed range of both $\Delta_{\sigma}$ and $\lambda_{\sigma\sigma\epsilon}$ for each choice of gap is summarized in Table~\ref{tab:spectrum}, as well as our estimates of the next several operators in the spectrum from the extremal functionals. Going to the nearly maximal (but still plausible) gap $\Delta_{\sigma'} = 2.875$, we obtain the fairly tight determinations $\Delta_{\sigma} = .58444(22)$, $\lambda_{\sigma\sigma\epsilon} = 1.0721(2)$, and $\lambda_{\epsilon\epsilon\epsilon} = 1.67(1)$. Lower gaps allow for a slightly wider range of these parameters, as can be read off of the table.

From the spectrum extraction, we see broad consistency with the range $\Delta_{\sigma'} = \Delta_{\epsilon'} - 1 = 2.86(4)$. For the third parity-odd operator $\sigma''$, we see an operator in the range $3.4(3)$ at lower values of $\Delta_{\sigma}$, but it appears to decouple at larger values and is replaced by a higher operator of dimension $5.6(4)$.  This decoupling merits further study, but it appears plausible to us that in the SCFT $\sigma''$ and the leading spin-$1$ operator $V_-$ live in a SUSY multiplet (where the lowest component is a fermion) and have dimension $\Delta_{\sigma''} = \Delta_{V_-} = 5.6(4)$. In addition, it is plausible that the second parity-even spin-$2$ operator $T'$ and leading parity-odd spin-$2$ operator $T^-$ live in a SUSY multiplet and have dimension $\Delta_{T'} = \Delta_{T^-} - 1 = 3.4(1)$. We also see the leading spin-$3$ operator in the range $\Delta_{3^-} = 4.6(2)$ and the leading spin-$4$ operator in the range $\Delta_{4^+} = 5.20(5)$. An estimate for the central charge can be read off the right bound of Fig.~\ref{fig:cT}, computed along the SUSY constraint curve with $\Delta_{\sigma'}^{\text{gap}} = 2.85$. We see that $C_T \gtrsim 1.45$ at the larger allowed values of $\Delta_{\sigma}$ and $C_T \gtrsim 1.52$ at the smaller values.

\begin{table}[t]
\centering
\resizebox{\textwidth}{!}{\begin{tabular}{|c|c|c|c|c|c|c|c|c|c|c|c|c|}
\hline
$\Delta_{\sigma'}^{\text{gap}}$ & $\Delta_\sigma = \Delta_{\epsilon} - 1$ & $\lambda_{\sigma \sigma \epsilon}$ & $\lambda_{\epsilon\epsilon\epsilon}$ & $\Delta_{\sigma'}$ & $\Delta_{\epsilon^{\prime}}$ & $\Delta_{\epsilon''}$ & $\Delta_{\sigma''}$  & $\Delta_{V^-}$ & $\Delta_{T'}$ & $\Delta_{T^-}$ & $\Delta_{3^-}$ & $\Delta_{4^+}$ \\
\specialrule{.1em}{.05em}{.05em}  
2.7 & min: 0.5839 & min: 1.0710 & 1.694 & 2.825 & 3.7 & 4.22 & 3.55 & 5.44 & 3.36 & 4.46 & 4.80 & 5.22 \\
\hline
2.7 & min: 0.5839 & max: 1.0714 & 1.695 & 2.835 & 3.7 & 4.16 & 3.75 & 5.54 & 3.20 & 4.46 & 4.81 & 5.17\\
 \hline
2.7 & max: 0.58478 & min: 1.0713 & 1.653 & 2.902 & 3.7 & 4.45 & 5.89 & 5.73 & 3.41 & 4.39 & 4.76 & 5.12 \\
\hline
2.7 & max: 0.58478 & max: 1.0717 & 1.654 & 2.906 & 3.7 & 4.37 & 6.01 & 5.96 & 3.34 & 4.38 & 4.79 & 5.23\\
\specialrule{.1em}{.05em}{.05em} 
2.75 & min: 0.584 & min: 1.0711 & 1.689 & 2.828 & 3.75 & 4.28 & 3.31 & 5.44 & 3.21 & 4.46 & 4.80 & 5.23 \\
\hline
2.75 & min: 0.584 & max: 1.0717 & 1.690 & 2.839 & 3.75 & 4.13 & 3.38 & 5.54 & 3.11 & 4.46 & 4.81 & 5.16\\
\hline
2.75 & max: 0.58478 & min: 1.0715 & 1.653 & 2.904 & 3.75 & 4.44 & 5.94 & 5.85 & 3.40 & 4.38 & 4.77 & 5.13\\
\hline
 2.75 & max: 0.58478 & max: 1.0717 & 1.654 & 2.906 & 3.75 & 4.46 & 6.00 & 5.96 & 3.37 & 4.37 & 4.79 & 5.22\\
\specialrule{.1em}{.05em}{.05em} 
2.8 & min: 0.58405 & min: 1.0714 & 1.688 & 2.846 & 3.8 & 4.31 & 3.47 & 5.51 & 3.27 & 4.46 & 4.79 & 5.24 \\
\hline
2.8 & min: 0.58405 & max: 1.0718 & 1.688 & 2.831 & 3.8 & 4.17 & 3.13 & 5.54 & 3.10 & 4.46 & 4.81 & 5.15\\
\hline
2.8 & max: 0.58473 & min: 1.0716 & 1.656 & 2.902 & 3.8 & 4.48 & 5.91 & 5.89 & 3.39 & 4.39 & 4.80 & 5.13\\
\hline
2.8 & max: 0.58473 & max: 1.0720 & 1.656 & 2.905 & 3.8 & 4.45 & 6.00 & 6.00 & 3.34 & 4.38 & 4.79 & 5.24 \\
\specialrule{.1em}{.05em}{.05em} 
2.875 & min: 0.58422 & min: 1.0719 & 1.680  & 2.875 & 3.875 & 4.42 & 5.36 & 5.07 & 3.29 & 4.46 & 4.80 & 5.21 \\
\hline
2.875 & min: 0.58422 & max: 1.0721 & 1.681  & 2.875 & 3.875 & 4.21 & 5.20 & 5.29 & 3.68 & 4.45 & 4.44 & 5.15 \\
\hline
2.875 & max: 0.58465 & min: 1.0720 & 1.660 & 2.901 & 3.875 & 4.55 & 5.91 & 6.01 & 3.35 & 4.39 & 4.58 & 5.25 \\
\hline
2.875 & max: 0.58465 & max: 1.0723 & 1.661 & 2.903 & 3.875 & 4.04 & 5.98 & 6.06 & 3.30 & 4.39 & 4.80 & 5.22 \\
\specialrule{.1em}{.05em}{.05em} 
\end{tabular}}
\caption{Extracted SUSY Ising CFT Spectra. For each gap $\Delta^{\text{gap}}_{\sigma'} = \Delta_{\epsilon'}^{\text{gap}}- 1$, we minimize and maximize the leading OPE coefficient $\lambda_{\sigma\sigma\epsilon}$ at each of the two intersections of the curves in Figure~\ref{fig:different_gaps} with the SUSY constraint given by Equation~\eqref{eq:susy}, corresponding to the minimum and maximum allowed values of $\Delta_{\sigma}$.
}
\label{tab:spectrum}
\end{table}

\newsec{Concluding with a Bold Conjecture}
\label{sec:conclusion}

The conformal bootstrap is one of the few fully nonperturbative tools we have for studying strongly-coupled quantum field theories. A dream of the bootstrap program, successfully carried out for minimal models in 2D, is to use bootstrap techniques to find exact analytical solutions to strongly-coupled systems. If we are optimistic, we can hope to take insights from the numerical bootstrap and use them to help us find such exact solutions in higher dimensions.

One of the most intriguing features of our analysis is the tendency for the closed supersymmetric islands in Figs.~\ref{fig:lambda_dep} and~\ref{fig:different_gaps} to be centered on and converging towards $\theta=1$, with our current best numerical window being $\theta = 1.000(3)$.\footnote{\hspace{1sp}\cite{Rong:2018okz} appears to improve this to $\theta=1.00009(39)$.} We are thus tempted to boldly conjecture the exact transcendental value 
\be
\frac{\lambda_{\epsilon\epsilon\epsilon}}{\lambda_{\sigma\sigma\epsilon}} &=& \tan(1).
\ee
Combining this conjecture with the SUSY relation in Eq.~(\ref{eq:susy}), we obtain the proposal
\be
\Delta_{\sigma} &=& \frac{15-2\tan(1) - \sqrt{4\tan(1)^2+36\tan(1)+9}}{18-8\tan(1)} \\
&=& .58445133696\ldots 
\ee
It would be exciting to check this possibility in more precise numerical bootstrap computations, perhaps involving mixed correlators with fermions~\cite{Iliesiu:2015akf} or the stress tensor~\cite{Dymarsky:2017yzx}, as well as to seek further analytical understanding of where such a relation might come from.

Now that we are getting a clearer picture of the low-lying spectrum of this theory, it will also be exciting to use analytical methods such as the lightcone bootstrap~\cite{Fitzpatrick:2012yx,Komargodski:2012ek,Alday:2015eya,Alday:2015ota,Alday:2015ewa,Alday:2016njk,Simmons-Duffin:2016wlq} and the inversion formula of~\cite{Caron-Huot:2017vep,Simmons-Duffin:2017nub} to obtain a more robust understanding of the leading-twist spectrum and eventually the full correlation functions of this theory. Such an understanding will help us learn how to compute RG flows in the vicinity of this fixed point and perhaps to try to make a connection to the non-supersymmetric lower kink identified in the fermion numerical bootstrap~\cite{Iliesiu:2015qra}, which is plausibly related to the non-supersymmetric GNY${}^*$ fixed point at $N=1$ identified in~\cite{Fei:2016sgs}. It will also be exciting to use bootstrap techniques to more fully study the interpolation between this theory and the 2D tri-critical Ising model, as can be done for the Ising model in fractional dimensions~\cite{El-Showk:2013nia}.

\ack{We would like to thank Soner Albayrak, Luca Iliesiu, Filip Kos, Daliang Li, David Meltzer, Silviu Pufu, Slava Rychkov, Michael Scherer, and David Simmons-Duffin for relevant discussions.  We thank the organizers of the Bootstrap 2018 workshop where a portion of this work was completed. This research is supported by NSF grant PHY-1350180 and Simons Foundation grant 488651 (Simons Collaboration on the Nonperturbative Bootstrap). The computations in this paper were run on the Grace computing cluster supported by the facilities and staff of the Yale University Faculty of Arts and Sciences High Performance Computing Center.}

\begin{appendices}
\section{Superconformal Three-Point Functions}
\label{app:3point}

We can obtain stronger constraints by incorporating relations between three-point function coefficients imposed by supersymmetry. Such constraints can be derived by constructing the allowed two- and three-point functions on superspace allowed by superconformal symmetry and expanding in the superspace coordinates. This formalism was developed for 3D $\mathcal{N}=1$ SCFTs by Park~\cite{Park:1999cw} and used more recently in the works~\cite{Nizami:2013tpa,Buchbinder:2015qsa}. The precise relations between super-descendant three-point functions needed for the bootstrap have not yet been fully worked out. In the present work we will derive and use the relation between $\lambda_{\sigma\sigma\epsilon}$ and $\lambda_{\epsilon\epsilon\epsilon}$ imposed by 3D $\mathcal{N}=1$ superconformal symmetry.

Concretely, we work with the supermultiplet $\Sigma(x,\theta) = \sigma(x) + \theta \psi(x) + \frac{i}{2} \theta^2 F(x)$. Using the formalism of~\cite{Park:1999cw} and following the spinor conventions of~\cite{Nizami:2013tpa}, the two-point function takes the form
\be
\<\Sigma(x_1, \theta_1) \Sigma(x_2, \theta_2)\> &=& \frac{1}{\tilde{x}_{12}^{2\Delta_{\sigma}}}
\ee
where $\tilde{x}_{12}^{\mu} = x_{12}^{\mu} + \frac{i}{2} \theta_1 \gamma^{\mu} \theta_2$ is the supertranslation invariant interval and $x_{12}^{\mu} = x_1^{\mu} - x_2^{\mu}$. 

Using $\tilde{x}_{12}^2 = x_{12}^2 + i \theta_1^{\alpha} (x_{12}^{\mu} \gamma_{\mu})_{\alpha}^{\beta} \theta_{2\beta} - \frac{3}{8} \theta_1^2 \theta_2^2$ and expanding to obtain the $\theta_1^2 \theta_2^2$ term yields
\be
\<F(x_1) F(x_2)\> &=& \Delta_{\sigma} (\Delta_{\sigma}-1/2) \frac{1}{x_{12}^{2(\Delta_{\sigma}+1)}},
\ee
so the operator $\epsilon(x) = \left[\Delta_{\sigma} (\Delta_{\sigma}-1/2)\right]^{-1/2} F(x)$ is canonically normalized.

Next we can consider the three-point function, which takes the form
\be
\<\Sigma(x_1,\theta_1) \Sigma(x_2,\theta_2) \Sigma(x_3,\theta_3)\> &=& \frac{H(X_{1},\Theta_1)}{\left(\tilde{x}_{12}^2\right)^{\Delta_{\sigma}}\left(\tilde{x}_{13}^2\right)^{\Delta_{\sigma}}}.
\ee
The function in the numerator must satisfy $H(X_{1}, \Theta_1) = \lambda^{\Delta_{\sigma}} H(\lambda X_1, \lambda^{1/2} \Theta_1)$, with
\be
X_{1} &=& X_{12-}^{-1} X_{23+} X_{31-}^{-1} = \frac{X_{12+} X_{23+} X_{31+}}{\tilde{x}_{12}^2 \tilde{x}_{13}^2}, \\
\Theta_{1\alpha} &=& i((X_{21+}^{-1} \theta_{21})_{\alpha} - (X_{31+}^{-1} \theta_{31})_{\alpha}).
\ee
Here $X_{12\pm} = \tilde{x}_{ij}^{\mu} \gamma_{\mu} \pm \frac{i}{4} \theta_{ij}^2 \mathds{1}$ and $\theta_{ij} = \theta_i - \theta_j$.

The only solution that is odd under the parity symmetry is
\be
H(X_1, \Theta_1) =  \lambda_- \frac{\Theta_1^2}{X_1^{\Delta_{\sigma}+1}}.
\ee

To help us expand this in components, we can use
\be
X_1^2 &=& \frac{\tilde{x}_{23}^2}{\tilde{x}_{12}^2 \tilde{x}_{13}^2}, \\
\Theta_1^2
&=& \frac{\theta_{21}^2}{\tilde{x}_{12}^2} + \frac{\theta_{31}^2}{\tilde{x}_{13}^2}   - 2 \frac{\theta_{21}^{\beta} (\eta_{\mu\nu}\delta_{\beta}^{\delta} +\epsilon_{\mu\nu\rho} (\gamma^{\rho})_{\beta}^{\delta}) \theta_{31\delta} \tilde{x}_{21}^{\mu} \tilde{x}_{31}^{\nu}}{\tilde{x}_{12}^2 \tilde{x}_{13}^2}.
\ee

Explicitly, the three-point structure is then
\be
\<\Sigma(x_1,\theta_1) \Sigma(x_2,\theta_2) \Sigma(x_3,\theta_3)\> 
&=& \lambda_- \frac{\tilde{x}_{13}^2\theta_{21}^2 + \tilde{x}_{12}^2 \theta_{31}^2   - 2 \theta_{21}^{\beta} (\eta_{\mu\nu}\delta_{\beta}^{\delta} +\epsilon_{\mu\nu\rho} (\gamma^{\rho})_{\beta}^{\delta}) \theta_{31\delta} \tilde{x}_{21}^{\mu} \tilde{x}_{31}^{\nu}}{\tilde{x}_{12}^{\Delta_{\sigma}+1}\tilde{x}_{13}^{\Delta_{\sigma}+1} \tilde{x}_{23}^{\Delta_{\sigma}+1}}.\nn\\
\ee
Expanding to order $\theta_3^2$, we immediately read off that 
\be
\<\sigma(x_1) \sigma(x_2) \epsilon(x_3)\> &=& \frac{\lambda_-}{\frac{i}{2}\left[\Delta_{\sigma} (\Delta_{\sigma}-1/2)\right]^{1/2} }\frac{1}{x_{12}^{\Delta_{\sigma}-1} x_{13}^{\Delta_{\sigma}+1} x_{23}^{\Delta_{\sigma}+1}},
\ee
so we see the relation
\be
\lambda_{\sigma\sigma\epsilon} &=&  \frac{\lambda_-}{\frac{i}{2}\left[\Delta_{\sigma} (\Delta_{\sigma}-1/2)\right]^{1/2} }.
\ee

Instead expanding to order $\theta_1^2 \theta_2^2 \theta_3^2$ gives (after some algebra)
\be
\<\epsilon(x_1) \epsilon(x_2) \epsilon(x_3)\> &=&  \frac{3(\Delta_{\sigma}-1)(3\Delta_{\sigma}-2)\lambda_{\sigma\sigma\epsilon}}{4\Delta_{\sigma}(\Delta_{\sigma}-1/2)}  \frac{1}{x_{12}^{\Delta_{\sigma}+1} x_{13}^{\Delta_{\sigma}+1} x_{23}^{\Delta_{\sigma}+1}},
 \ee
 so we conclude that
 \be
\frac{\lambda_{\epsilon\epsilon\epsilon}}{\lambda_{\sigma\sigma\epsilon}} &=& \frac{3(\Delta_{\sigma}-1)(3\Delta_{\sigma}-2)}{4\Delta_{\sigma}(\Delta_{\sigma}-1/2)}.
 \ee
 
 \section{SDPB Parameters}
 \label{app:SDPB}
When passing the problem to SDPB~\cite{Simmons-Duffin:2015qma}, in most plots we used a primal error threshold of $10^{-120}$ and a dual error threshold of $10^{-10}$, together with a precision of 600 binary digits. At each value of the derivative order $\Lambda$, we adopt a different cutoff on spin $\ell_{\max}$ and number of conformal block poles $\nu$. The conformal blocks are computed using cboot~\cite{CBoot}. Additional Python scripts used in this project are available at~\cite{pythonscripts}. The parameters we used are shown in the table below. 

Note that at $\Lambda=25$, most plots used $\nu=21$, but in generating Figures \ref{fig:susy_line_regions}, \ref{fig:cT}, and Table \ref{tab:spectrum}, the higher value of $\nu=27$ was used in order to ensure the stability of the spectrum extraction. These computations additionally used a precision of 1000 binary digits and a value of $10^{-30}$ for both primal and dual error thresholds. The extraction itself is done with the python script at~\cite{spectrum}, using a zero threshold of $10^{-20}$. 
\vspace{.5cm}
 \begin{table}[!h]
	\begin{center}
		\begin{tabular}{|c|c|c|c|}
			\hline
			$\Lambda$ & $\ell_{max}$ & $\nu$ \\
			\hline \hline
			$9$ & $20$ & $12$ \\
			$13$ & $29$ & $14$ \\
			$17$ & $51$ & $19$ \\
			$21$ & $51$ & $20$ \\
			$25$ & $51$ & $21, 27$\\
			\hline
		\end{tabular}
	\end{center}
    \vspace{-12pt}
\end{table} 
 
 \end{appendices}

\newpage
\bibliography{references}

\begin{thebibliography}{10}
\ifx\href\asklfhas\newcommand{\href}[2]{#2}\fi
\ifx\arxivref\asklfhas\newcommand{\arxivref}[2]{\href{http://arxiv.org/abs/#1}{#2}}\fi
\ifx\doiref\asklfhas\newcommand{\doiref}[2]{\href{http://dx.doi.org/#1}{#2}}\fi
\parskip 0pt
\normalsize

\bibitem{ElShowk:2012ht}
S.~El-Showk, M.~F. Paulos, D.~Poland, S.~Rychkov, D.~Simmons-Duffin \&
  A.~Vichi,
\textit{``{Solving the 3D Ising Model with the Conformal Bootstrap}''},
\doiref{10.1103/PhysRevD.86.025022}{Phys.Rev. \textbf{D86}, 025022 (2012)},
\normalsize{\texttt{\arxivref{1203.6064}{arXiv:1203.6064}}}

\bibitem{Kos:2013tga}
F.~Kos, D.~Poland \& D.~Simmons-Duffin,
\textit{``{Bootstrapping the $\text{O}(N)$ vector models}''},
\doiref{10.1007/JHEP06(2014)091}{JHEP \textbf{1406}, 091 (2014)},
\normalsize{\texttt{\arxivref{1307.6856}{arXiv:1307.6856}}}

\bibitem{El-Showk:2014dwa}
S.~El-Showk, M.~F. Paulos, D.~Poland, S.~Rychkov, D.~Simmons-Duffin et~al.,
\textit{``{Solving the 3d Ising Model with the Conformal Bootstrap II.
  c-Minimization and Precise Critical Exponents}''},
\doiref{10.1007/s10955-014-1042-7}{J.Stat.Phys. \textbf{157}, 869 (2014)},
\normalsize{\texttt{\arxivref{1403.4545}{arXiv:1403.4545}}}

\bibitem{Kos:2014bka}
F.~Kos, D.~Poland \& D.~Simmons-Duffin,
\textit{``{Bootstrapping Mixed Correlators in the 3D Ising Model}''},
\doiref{10.1007/JHEP11(2014)109}{JHEP \textbf{1411}, 109 (2014)},
\normalsize{\texttt{\arxivref{1406.4858}{arXiv:1406.4858}}}

\bibitem{Kos:2015mba}
F.~Kos, D.~Poland, D.~Simmons-Duffin \& A.~Vichi,
\textit{``{Bootstrapping the O(N) Archipelago}''},
\doiref{10.1007/JHEP11(2015)106}{JHEP \textbf{1511}, 106 (2015)},
\normalsize{\texttt{\arxivref{1504.07997}{arXiv:1504.07997}}}

\bibitem{Kos:2016ysd}
F.~Kos, D.~Poland, D.~Simmons-Duffin \& A.~Vichi,
\textit{``{Precision Islands in the Ising and $O(N)$ Models}''},
\doiref{10.1007/JHEP08(2016)036}{JHEP \textbf{1608}, 036 (2016)},
\normalsize{\texttt{\arxivref{1603.04436}{arXiv:1603.04436}}}

\bibitem{Poland:2016chs}
D.~Poland \& D.~Simmons-Duffin,
\textit{``{The conformal bootstrap}''},
\doiref{10.1038/nphys3761}{Nature~Phys. \textbf{12}, 535 (2016)}

\bibitem{Poland:2018epd}
D.~Poland, S.~Rychkov \& A.~Vichi,
\textit{``{The Conformal Bootstrap: Theory, Numerical Techniques, and
  Applications}''},
\normalsize{\texttt{\arxivref{1805.04405}{arXiv:1805.04405}}}

\bibitem{Iliesiu:2015qra}
L.~Iliesiu, F.~Kos, D.~Poland, S.~S. Pufu, D.~Simmons-Duffin \& R.~Yacoby,
\textit{``{Bootstrapping 3D Fermions}''},
\doiref{10.1007/JHEP03(2016)120}{JHEP \textbf{1603}, 120 (2016)},
\normalsize{\texttt{\arxivref{1508.00012}{arXiv:1508.00012}}}

\bibitem{Iliesiu:2017nrv}
L.~Iliesiu, F.~Kos, D.~Poland, S.~S. Pufu \& D.~Simmons-Duffin,
\textit{``{Bootstrapping 3D Fermions with Global Symmetries}''},
\doiref{10.1007/JHEP01(2018)036}{JHEP \textbf{1801}, 036 (2018)},
\normalsize{\texttt{\arxivref{1705.03484}{arXiv:1705.03484}}}

\bibitem{Grover:2013rc}
T.~Grover, D.~N. Sheng \& A.~Vishwanath,
\textit{``{Emergent Space-Time Supersymmetry at the Boundary of a Topological
  Phase}''},
\doiref{10.1126/science.1248253}{Science \textbf{344}, 280 (2014)},
\normalsize{\texttt{\arxivref{1301.7449}{arXiv:1301.7449}}}

\bibitem{Fei:2016sgs}
L.~Fei, S.~Giombi, I.~R. Klebanov \& G.~Tarnopolsky,
\textit{``{Yukawa CFTs and Emergent Supersymmetry}''},
\doiref{10.1093/ptep/ptw120}{PTEP \textbf{2016}, 12C105 (2016)},
\normalsize{\texttt{\arxivref{1607.05316}{arXiv:1607.05316}}}

\bibitem{Zerf:2017zqi}
N.~Zerf, L.~N. Mihaila, P.~Marquard, I.~F. Herbut \& M.~M. Scherer,
\textit{``{Four-loop critical exponents for the Gross-Neveu-Yukawa models}''},
\doiref{10.1103/PhysRevD.96.096010}{Phys.~Rev. \textbf{D96}, 096010 (2017)},
\normalsize{\texttt{\arxivref{1709.05057}{arXiv:1709.05057}}}

\bibitem{Ihrig:2018hho}
B.~Ihrig, L.~N. Mihaila \& M.~M. Scherer,
\textit{``{Critical behavior of Dirac fermions from perturbative
  renormalization}''},
\normalsize{\texttt{\arxivref{1806.04977}{arXiv:1806.04977}}}

\bibitem{Park:1999cw}
J.-H. Park,
\textit{``{Superconformal symmetry in three-dimensions}''},
\doiref{10.1063/1.1290056}{J.~Math.~Phys. \textbf{41}, 7129 (2000)},
\normalsize{\texttt{\arxivref{hep-th/9910199}{hep-th/9910199}}}

\bibitem{Rong:2018okz}
J.~Rong \& N.~Su,
\textit{``{Bootstrapping minimal $\mathcal{N}=1$ superconformal field theory in
  three dimensions}''},
\normalsize{\texttt{\arxivref{1807.04434}{arXiv:1807.04434}}}

\bibitem{Simmons-Duffin:2015qma}
D.~Simmons-Duffin,
\textit{``{A Semidefinite Program Solver for the Conformal Bootstrap}''},
\doiref{10.1007/JHEP06(2015)174}{JHEP \textbf{1506}, 174 (2015)},
\normalsize{\texttt{\arxivref{1502.02033}{arXiv:1502.02033}}}

\bibitem{CBoot}
T.~Ohtsuki,
\textit{``{CBoot: A sage module to create (convolved) conformal block
  table}''},
\href{{https://github.com/tohtsky/cboot}}{\texttt{{https://github.com/tohtsky/cboot}}}

\bibitem{Bashkirov:2013vya}
D.~Bashkirov,
\textit{``{Bootstrapping the $\mathcal{N} = 1$ SCFT in three dimensions}''},
\normalsize{\texttt{\arxivref{1310.8255}{arXiv:1310.8255}}}

\bibitem{Li:2017kck}
Z.~Li \& N.~Su,
\textit{``{3D CFT Archipelago from Single Correlator Bootstrap}''},
\normalsize{\texttt{\arxivref{1706.06960}{arXiv:1706.06960}}}

\bibitem{Poland:2010wg}
D.~Poland \& D.~Simmons-Duffin,
\textit{``{Bounds on 4D Conformal and Superconformal Field Theories}''},
\doiref{10.1007/JHEP05(2011)017}{JHEP \textbf{1105}, 017 (2011)},
\normalsize{\texttt{\arxivref{1009.2087}{arXiv:1009.2087}}}

\bibitem{ElShowk:2012hu}
S.~El-Showk \& M.~F. Paulos,
\textit{``{Bootstrapping Conformal Field Theories with the Extremal Functional
  Method}''},
\doiref{10.1103/PhysRevLett.111.241601}{Phys.~Rev.~Lett. \textbf{111}, 241601
  (2013)},
\normalsize{\texttt{\arxivref{1211.2810}{arXiv:1211.2810}}}

\bibitem{Iliesiu:2015akf}
L.~Iliesiu, F.~Kos, D.~Poland, S.~S. Pufu, D.~Simmons-Duffin \& R.~Yacoby,
\textit{``{Fermion-Scalar Conformal Blocks}''},
\doiref{10.1007/JHEP04(2016)074}{JHEP \textbf{1604}, 074 (2016)},
\normalsize{\texttt{\arxivref{1511.01497}{arXiv:1511.01497}}}

\bibitem{Dymarsky:2017yzx}
A.~Dymarsky, F.~Kos, P.~Kravchuk, D.~Poland \& D.~Simmons-Duffin,
\textit{``{The 3d Stress-Tensor Bootstrap}''},
\doiref{10.1007/JHEP02(2018)164}{JHEP \textbf{1802}, 164 (2018)},
\normalsize{\texttt{\arxivref{1708.05718}{arXiv:1708.05718}}}

\bibitem{Fitzpatrick:2012yx}
A.~L. Fitzpatrick, J.~Kaplan, D.~Poland \& D.~Simmons-Duffin,
\textit{``{The Analytic Bootstrap and AdS Superhorizon Locality}''},
\doiref{10.1007/JHEP12(2013)004}{JHEP \textbf{1312}, 004 (2013)},
\normalsize{\texttt{\arxivref{1212.3616}{arXiv:1212.3616}}}

\bibitem{Komargodski:2012ek}
Z.~Komargodski \& A.~Zhiboedov,
\textit{``{Convexity and Liberation at Large Spin}''},
\doiref{10.1007/JHEP11(2013)140}{JHEP \textbf{1311}, 140 (2013)},
\normalsize{\texttt{\arxivref{1212.4103}{arXiv:1212.4103}}}

\bibitem{Alday:2015eya}
L.~F. Alday, A.~Bissi \& T.~Lukowski,
\textit{``{Large spin systematics in CFT}''},
\doiref{10.1007/JHEP11(2015)101}{JHEP \textbf{1511}, 101 (2015)},
\normalsize{\texttt{\arxivref{1502.07707}{arXiv:1502.07707}}}

\bibitem{Alday:2015ota}
L.~F. Alday \& A.~Zhiboedov,
\textit{``{Conformal Bootstrap With Slightly Broken Higher Spin Symmetry}''},
\doiref{10.1007/JHEP06(2016)091}{JHEP \textbf{1606}, 091 (2016)},
\normalsize{\texttt{\arxivref{1506.04659}{arXiv:1506.04659}}}

\bibitem{Alday:2015ewa}
L.~F. Alday \& A.~Zhiboedov,
\textit{``{An Algebraic Approach to the Analytic Bootstrap}''},
\doiref{10.1007/JHEP04(2017)157}{JHEP \textbf{1704}, 157 (2017)},
\normalsize{\texttt{\arxivref{1510.08091}{arXiv:1510.08091}}}

\bibitem{Alday:2016njk}
L.~F. Alday,
\textit{``{Large Spin Perturbation Theory for Conformal Field Theories}''},
\doiref{10.1103/PhysRevLett.119.111601}{Phys.~Rev.~Lett. \textbf{119}, 111601
  (2017)},
\normalsize{\texttt{\arxivref{1611.01500}{arXiv:1611.01500}}}

\bibitem{Simmons-Duffin:2016wlq}
D.~Simmons-Duffin,
\textit{``{The Lightcone Bootstrap and the Spectrum of the 3d Ising CFT}''},
\doiref{10.1007/JHEP03(2017)086}{JHEP \textbf{1703}, 086 (2017)},
\normalsize{\texttt{\arxivref{1612.08471}{arXiv:1612.08471}}}

\bibitem{Caron-Huot:2017vep}
S.~Caron-Huot,
\textit{``{Analyticity in Spin in Conformal Theories}''},
\doiref{10.1007/JHEP09(2017)078}{JHEP \textbf{1709}, 078 (2017)},
\normalsize{\texttt{\arxivref{1703.00278}{arXiv:1703.00278}}}

\bibitem{Simmons-Duffin:2017nub}
D.~Simmons-Duffin, D.~Stanford \& E.~Witten,
\textit{``{A spacetime derivation of the Lorentzian OPE inversion formula}''},
\doiref{10.1007/JHEP07(2018)085}{JHEP \textbf{1807}, 085 (2018)},
\normalsize{\texttt{\arxivref{1711.03816}{arXiv:1711.03816}}}

\bibitem{El-Showk:2013nia}
S.~El-Showk, M.~Paulos, D.~Poland, S.~Rychkov, D.~Simmons-Duffin \& A.~Vichi,
\textit{``{Conformal Field Theories in Fractional Dimensions}''},
\doiref{10.1103/PhysRevLett.112.141601}{Phys.~Rev.~Lett. \textbf{112}, 141601
  (2014)},
\normalsize{\texttt{\arxivref{1309.5089}{arXiv:1309.5089}}}

\bibitem{Nizami:2013tpa}
A.~A. Nizami, T.~Sharma \& V.~Umesh,
\textit{``{Superspace formulation and correlation functions of 3d
  superconformal field theories}''},
\doiref{10.1007/JHEP07(2014)022}{JHEP \textbf{1407}, 022 (2014)},
\normalsize{\texttt{\arxivref{1308.4778}{arXiv:1308.4778}}}

\bibitem{Buchbinder:2015qsa}
E.~I. Buchbinder, S.~M. Kuzenko \& I.~B. Samsonov,
\textit{``{Superconformal field theory in three dimensions: Correlation
  functions of conserved currents}''},
\doiref{10.1007/JHEP06(2015)138}{JHEP \textbf{1506}, 138 (2015)},
\normalsize{\texttt{\arxivref{1503.04961}{arXiv:1503.04961}}}

\bibitem{pythonscripts}
A.~Atanasov, A.~Hillman \& D.~Poland,
\href{{https://github.com/ABAtanasov/IsingBootstrap/}}{\texttt{{https://github.com/ABAtanasov/IsingBootstrap}}}

\bibitem{spectrum}
D.~Simmons-Duffin,
\href{{https://gitlab.com/bootstrapcollaboration/spectrum-extraction}}{\texttt{{https://gitlab.com/bootstrapcollaboration/spectrum-extraction}}}

\end{thebibliography}

\end{document}